%% file: main.tex
\documentclass[conference, rgb]{IEEEtran}
\IEEEoverridecommandlockouts

\usepackage{cite}
\usepackage{amsmath,amssymb,amsthm}
\usepackage{algorithmic}
\usepackage{graphicx}
\usepackage{textcomp}
\usepackage{xcolor}
\usepackage{tikz}
\usepackage{pgfplots}
\pgfplotsset{compat=1.17}
\usepackage{hyperref}
\usepackage{pifont}
\usepackage{multirow}
\newcommand{\cmark}{\ding{51}}%
\newcommand{\xmark}{\ding{55}}%

\newcommand{\SU}{\operatorname{SU}}

\begin{document}

\title{Analysis of quantum neural network performance via edge cases
}

\author{\IEEEauthorblockN{Maximilian Balthasar Mansky\IEEEauthorrefmark{1}\IEEEauthorrefmark{2}, Tobias Rohe\IEEEauthorrefmark{1}, Dmytro Bondarenko\IEEEauthorrefmark{1},\\ Linus Menzel\IEEEauthorrefmark{1} and Claudia Linnhoff-Popien\IEEEauthorrefmark{1}}
\IEEEauthorblockA{\IEEEauthorrefmark{1}Institute of Informatics, LMU Munich}
\IEEEauthorblockA{\IEEEauthorrefmark{2}
Email: maximilian-balthasar.mansky@ifi.lmu.de}}

\maketitle

\begin{abstract}
We evaluate the particular performance of different quantum machine learning networks on a graph classification task. Quantum circuits with varying internal symmetry that completely, partially and not at all confer to the symmetry of the graph show different performance on the data set. The convergence results are inspected using a number of special graphs with particular structure. These are unlikely to occur in the training data and cover specific cases that refute the assumption that the quantum neural network learns simpler surrogate models based on the number of edges in the graph.
\end{abstract}

\begin{IEEEkeywords}
    quantum machine learning, symmetry, discrete symmetry, optimization, graph problems, edge case analysis
\end{IEEEkeywords}

\section{Introduction}

The analysis of neural networks, both quantum and classical, is difficult for high-dimensional data sets. In simple two-dimensional cases, examining the separation boundaries allows some insight into the structure of the network. For supervised learning with features and labels, the decision boundaries for the different labels can be drawn in the feature data. In the oft-used two-moons dataset, it can be used to check for erroneous classifications or islands of one class within another. As the dimensionality of the data set increases, this approach quickly becomes unavailable due to the difficulty of mapping data and separation boundaries down to visually examinable dimensions.

However, there are classes of data for which an analysis is still possible. One such class of data contains graphs. Graphs are high-dimensional and complex objects that can still be succinctly represented in two dimensions. In this work, we use quantum neural networks to classify graphs according to a global property, connectedness. In another work, we have presented results that indicate that specially constructed quantum neural networks with permutation invariance \cite{mansky_solving_2025} perform well on this task, compared to standard approaches. One point of contention is that connectedness is strongly related to the number of edges in the graph \cite{erdos_random_1959}. For random Erdős-Rényi graphs, the probability for a graph to be connected quickly approaches one as the edge probability $p_\text{edge}$ approaches $p_\text{edge}\to \ln(n)/n$. It is therefore possible that the neural network simply 'learns' to count the number of edges and approximate the threshold \cite{erdos_random_1959}. 

This kind of surrogate learning \cite{cozad_learning_2014} is difficult to detect using typical approaches of viewing accuracy plots or ROC curves \cite{silhavy_review_2023}. In the present case, we can evaluate the structure of the network in more detail since individual data points can be mapped to visually accessible dimensions. While this approach is not available for all possible data sets, in the present case it is an important tool to refute some of the possible surrogate models.

\section{Related work}

The evaluation of machine learning models is an important topic, yet it is often relegated to individual cases. The inherent focus is often on metrics that capture the overall performance of the model \cite{japkowicz_performance_2015, colliot_evaluating_2023, dwivedi_performance_2018}. Different ansatzes have been proposed \cite{jiao_performance_2016, silhavy_review_2023}, up to a measurement theory for capturing the performance of networks \cite{flach_performance_2019}. Appropriate evaluation mechanisms often depend on the experience of the researcher and the training task at hand.

Surrogate models are not detrimental per se and find application in the simplification and approximation of large models \cite{kim_machine_2020, trinchero_machine_2018, cozad_learning_2014}. The surrogate model is then designed to capture the essential structure of the larger model or expensive simulation for a faster exploration of the model and feature space.

\section{Methods}

\begin{figure}[htb]
    \centering
    \begin{tikzpicture}
        \begin{axis}[small,
height=5cm, width=9cm,
no markers,
xlabel = {Graph edge probability $p$},
axis x line = bottom,
axis y line = left,
ymajorgrids,
major grid style = {very thin, gray!50},
major tick style = {very thin, gray!50},
axis line style={gray},
axis line shift=2pt,
xmin = 0,
xmax = 1.01,
ylabel = {Prob. connectedness},
every axis y label/.style={at={(ticklabel cs:.5)},rotate = 90, anchor=center},
ymax = 1.015,
title = {Probability of connectedness for 8 nodes},
	]
		\addplot [red] table [x=p, y=connected, col sep=comma]{data/connected-probability.csv};
\end{axis}
\end{tikzpicture}
    \caption{The relationship between the edge probability and the likelihood of graph connectedness for eight nodes in the Erdős-Rényi random graph model.}
    \label{fig:connectedness-likelihood}
\end{figure}
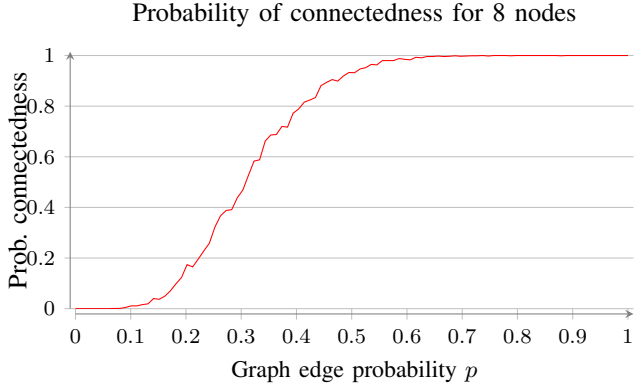

\paragraph{Hypothesis} This hypothesis can be examined using an analysis on particular edge cases. For example, the property can be tested on an $n$-node graph with a complete subgraph of $n-1$ nodes and one disconnected node. The number of edges is very high, yet the graph is disconnected. In the Erdős-Rényi model, a graph like this is extremely unlikely to occur, as the edge probability between each pair of nodes is independent from each other. This relationship between edge probability and connectedness probability is shown in figure \ref{fig:connectedness-likelihood}. As the number of nodes $n$ increases, the transition sharpens.

\paragraph{Data set}
We train the quantum neural network with a dataset of random graphs that are classified according to their property. Each epoch of training consists of 100 examples of graphs, with a 50\% split to each category of connected and unconnected graph. The performance is measured on a validation set of 2900 graphs, also evenly split. Due to the limitations of simulation, the graphs have eight nodes. Each node is mapped to a qubit, with edges drawn as CZ gates. CZ gates are abelian and self-inverse, mirroring the properties of unweighted graph edges. The edge encoding is sandwiched with Hadamard gates, for a graph encoding layer defined as:
\begin{equation}
    |\mathcal{G}\rangle = \bigotimes_{i=1}^n \text{H} \bigotimes_{i,j \in E}\text{CZ}_{ij} \bigotimes_{i=1}^n\text{H}|0\rangle^{\otimes n} \label{eq:embedding}
\end{equation}

\paragraph{Quantum circuits}
The quantum neural network consists of layers that are constructed according to their desired property. The permutation-invariant quantum circuit is defined through a permutation symmetry on the qubits. That is, it is invariant under a transformation $\text{SWAP}_{ij}U_\text{pi}\text{SWAP}_{ij}$. The details of this construction are given in \cite{mansky_permutation-invariant_2023, mansky_scaling_2024}. The layers of the quantum circuit are shown in table \ref{tab:circuits}. The layers are concatenated until approximately the same number of parameters is reached, for a goal of 120 parameters in each of the three different quantum circuits. Since the parameters are shared across the layers, this means 40 layers in the case of the permutation-invariant quantum circuit, 30 for the cyclic-invariant quantum circuit and 3 for the strongly entangling layer from pennylane \cite{bergholmPennyLaneAutomaticDifferentiation2022}, here termed the ``standard ansatz''. The measurement that determines the label is a Z$^n$ measurement on all qubits that returns a single number between $-1$ and $+1$. Negative numbers are mapped to the `disconnected' label and positive numbers to `connected'. The measurement is permutation invariant in the same sense as above.

\begin{table}
\caption[Quantum circuit building blocks]{The quantum circuits building blocks used in the quantum machine learning analysis. Each quantum circuit diagram shows one layer of the quantum circuit. This layer is repeated until the desired number of parameters is achieved. Gates with the same color contain shared parameters. $U$ gates indicate gates  that cover the whole $\SU(2)$ sphere. From top to bottom: Permutation-invariant, cyclic invariant and strongly-entangling layer.}\label{tab:circuits}
\centering

\begin{tikzpicture}[gate/.style={rectangle, draw=black, fill=white, inner sep=3pt}, xscale=.53, yscale=.45, baseline=(current bounding box.center)]
\path (2,1) -- (2, 7);
\foreach \i in {1,2,...,6} {
\draw (0.5, \i) --+ (17, 0);
\draw (1, \i) node[gate, fill=red!20] {X}
	(2, \i) node[gate, fill=orange!20] {Y};
	\foreach \j in {3, 4, ..., 8} {
	\node (p\j\i) at (\j, \i) {\phantom{$ZZ$}};};};
\foreach \i in {1,2, ..., 5} {
\draw[line width=.3pt, line cap=rect, double=purple!20, double distance=12pt] (\i + 2, 1) --+ (0, \i);
\draw (\i + 2, 1) node {ZZ};};

\foreach \i in {1,2, ..., 4} {
\draw[line width=.3pt, line cap=rect, double=purple!20, double distance=12pt] (\i + 7, 2) --+ (0, \i);
\draw (\i + 7, 2) node {ZZ};
\draw (\i + 2.5, \i + 1) --+ (5 - \i, 0);};

\foreach \i in {1,2, ..., 3} {
\draw[line width=.3pt, line cap=rect, double=purple!20, double distance=12pt] (\i + 11, 3) --+ (0, \i);
\draw (\i + 11, 3) node {ZZ};
\draw (\i + 7.5, \i + 2) --+ (4 - \i, 0);};

\foreach \i in {1,2} {
\draw[line width=.3pt, line cap=rect, double=purple!20, double distance=12pt] (\i + 14, 4) --+ (0, \i);
\draw (\i + 14, 4) node {ZZ};
\draw (\i + 11.5, \i + 3) --+ (3 - \i, 0);};

\foreach \i in {1} {
\draw[line width=.3pt, line cap=rect, double=purple!20, double distance=12pt] (\i + 16, 5) --+ (0, \i);
\draw (\i + 16, 5) node {ZZ};
\draw (\i + 14.5, \i + 4) --+ (2 - \i, 0);};
\end{tikzpicture}

\begin{tikzpicture}[gate/.style={rectangle, draw=black, fill=white, inner sep=3pt}, xscale=.6, yscale=.45, baseline=(current bounding box.center)]
\path (2,1) -- (2, 7);
\foreach \i in {1,2,...,6} {
\draw (0.5, \i) --+ (14.0, 0);
\draw (1, \i) node[gate, fill=red!20] {X}
	(2, \i) node[gate, fill=orange!20] {Y};
	};
\foreach \i in {1,2, ..., 5} {
\draw[line width=.3pt, line cap=rect, double=purple!20, double distance=14pt] (\i + 2, \i) --+ (0, 1);
\draw (\i + 2, \i) node {ZZ};};
\draw[line width=.3pt, line cap=rect, double=purple!20, double distance=14pt] (8, 1) --+ (0, 5);
\draw (8,1) node {ZZ};
\foreach \i in {2, 3, ..., 5} {
\draw (7.5, \i) --+(1,0);};
\foreach \i in {1,2, ..., 4} {
\draw[line width=.3pt, line cap=rect, double=violet!20, double distance=14pt] (\i + 8.0, \i) --+ (0, 2);
\draw (\i + 7.5, \i+1) --+ (1,0);
\draw (\i + 8.0, \i) node {ZZ};};
\foreach \i in {1,2} {
\draw[line width=.3pt, line cap=rect, double=violet!20, double distance=14pt] (12.0 + \i, \i) --+ (0, 4);
	\foreach \j in {1, 2, 3} {
	\draw (11.5 + \i, \i + \j) --+ (1,0);};
\draw (12.0 + \i,\i) node {ZZ};};
\end{tikzpicture}

\begin{tikzpicture}[gate/.style={rectangle, draw=black, fill=white, inner sep=3pt}, xscale=.6, yscale=.45, baseline=(current bounding box.center)]
\path (2,1) -- (2, 7);
\begin{scope}[yscale=-1, yshift=-7cm]
\foreach \i/\col in {1/0, 2/20, 3/40, 4/60, 5/80, 6/100} {
\draw (0.5, \i) --+ (14, 0);
\draw (1, \i) node[gate, fill=red!\col!violet!20] {U};
	};
\foreach \i in {5,4, ..., 1} {
\path (\i + 1, \i+1) node[circle, draw=black] (target) {} (\i + 1, \i) node[circle, fill=black, inner sep=1.3pt] (control) {};
\draw (target.south) -- (control.center);};
\path (7, 1) node[circle, draw=black] (target) {} (7, 6) node[circle, fill=black, inner sep=1.3pt] (control) {};
\draw(target.north) -- (control.center);

\begin{scope}[xshift = 7cm]
\foreach \i/\col in {1/0, 2/20, 3/40, 4/60, 5/80, 6/100} {
\draw (1, \i) node[gate, fill=orange!\col!purple!20] {U};
	};
\foreach \i in {1,2, ..., 4} {
\path (\i + 1, \i+2) node[circle, draw=black] (target) {} (\i + 1, \i) node[circle, fill=black, inner sep=1.3pt] (control) {};
\draw (target.south) -- (control.center);};
\foreach \i in {1,2} {
\path (5 + \i, \i) node[circle, draw=black] (target) {} (5 + \i, 4 + \i) node[circle, fill=black, inner sep=1.3pt] (control) {};
\draw(target.north) -- (control.center);};
\end{scope}
\end{scope}

\end{tikzpicture}

\end{table}

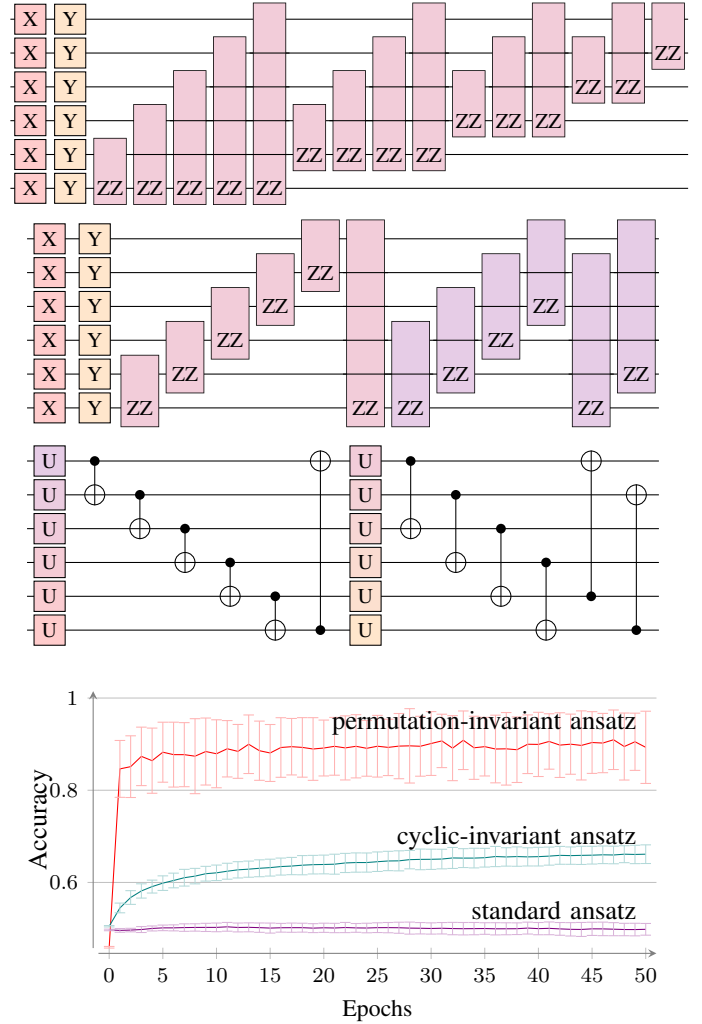
\begin{figure}[htb]
    \centering
    \begin{tikzpicture}[baseline=(current bounding box.center)]
\begin{axis}[small,
height=5cm, width=9cm,
no markers,
xlabel = {Epochs},
ylabel = {Accuracy},
axis x line = bottom,
axis y line = left,
ymajorgrids,
major grid style = {very thin, gray!50},
major tick style = {very thin, gray!50},
axis line style={gray},
axis line shift=2pt,
xmin = -1,
xmax = 51,
every axis y label/.style={at={(ticklabel cs:.5)},rotate = 90, anchor=center},
ymax = 1.015,
	]
		\addplot [red, error bars/.cd, y dir = both, y explicit, error bar style={ thin, red!30}] table [x=epochs, y=Sn, col sep=comma, y error = sn-error]{data/graph-connectedness-8.csv};
	\addplot [teal, error bars/.cd, y dir=both, y explicit, error bar style={very thin, teal!30}] table [x=epochs, y=Cn2, col sep=comma, y error = cn2-error]{data/graph-connectedness-8.csv};
	\addplot [violet, error bars/.cd, y dir=both, y explicit, error bar style={very thin, violet!30}] table [x=epochs, y=entanglement, col sep=comma, y error=en-error]{data/graph-connectedness-8.csv};
	\draw[red] (yticklabel cs: 0) -- (yticklabel cs: 1);
	\draw (axis cs: 50, .9) node[above left] {permutation-invariant ansatz}
		(axis cs: 50, .65) node[above left] {cyclic-invariant ansatz}
		(axis cs: 50, .50) node[above left] {standard ansatz};

\end{axis}
\end{tikzpicture}
    \caption{The performance of the different quantum circuits through the training epochs on the data set. The error bars indicate the $3\sigma$ deviation for the 10 simulation runs for each quantum circuit.}
    \label{fig:connectedness-result}
\end{figure}

\paragraph{Training results}
The results of the training are presented in figure \ref{fig:connectedness-result}. The choice of symmetry makes a significant difference in the performance of the network. The binary classification problem of connectedness is permutation-invariant, in the sense that the order of the nodes does not matter. This is captured by the permutation-invariant neural network and to a lesser extend by the cyclic-invariant quantum circuit. A detailed discussion of the results can be found in \cite{mansky_solving_2025}.

\section{Edge case testing}

With the quantum circuits and their results on the data set established, it is now possible to examine the performance of the models in detail. To this end, we examine their performance on particular graphs. The graphs and their classification results are shown in table \ref{tab:edge-case-results}. 

\begin{table}[htb]
    \centering
        \caption{The results for the different test graphs under classification by the different quantum circuits. Negative measurement indicate disconnected, positive values connected graphs. The minimum and maximum values are $-1$ and $+1$ respectively.}
        \normalsize
\begin{tabular}{c|l l c}
Graph & Circuit & Result & Correct\\\hline
\multirow{3}*{
    \begin{tikzpicture}[scale=0.5]
    \foreach \i in {1, ..., 8} {
    	\draw (\i * 360 / 8:1) node[circle, fill=black, inner sep = 1pt] (n\i) {};
    };
    \foreach \i in {1, ..., 7} {
    	\foreach \j in {1, ..., \i} {
    		\draw (n\i) -- (n\j);
    	};
    };
    \end{tikzpicture}
}	& perm-inv & $-0.4$ & \cmark\\
	& cyc-inv & $-4 \times 10^{-4}$ & \cmark\\
	& strongly-e & $0.08$ & \xmark\\\hline
	
\multirow{3}*{
    \begin{tikzpicture}[scale=0.5]
    \foreach \i in {1, ..., 8} {
    	\draw (\i * 360 / 8:1) node[circle, fill=black, inner sep = 1pt] (n\i) {};
    };
    \foreach \i in {1, ..., 7} {
    	\foreach \j in {1, ..., \i} {
    		\draw (n\i) -- (n\j);
    	};
    };
    \draw (n7) -- (n8);
    \end{tikzpicture}
}	& perm-inv & $-0.3$ & \xmark\\
	& cyc-inv & $-6 \times 10^{-3}$ & \xmark\\
	& strongly-e & $3 \times 10^{-3}$ & \cmark\\\hline
	
\multirow{3}*{
    \begin{tikzpicture}[scale=0.5]
    \foreach \i in {1, ..., 8} {
    	\draw (\i * 360 / 8:1) node[circle, fill=black, inner sep = 1pt] (n\i) {};
    };
    \foreach \i in {1, ..., 4} {
    	\foreach \j in {1, ..., \i} {
    		\draw (n\i) -- (n\j);
    	};
    };
    \foreach \i in {5, ..., 8} {
    	\foreach \j in {5, ..., \i} {
    		\draw (n\i) -- (n\j);
    	};
    };
    \end{tikzpicture}
}	& perm-inv & $0.2$ & \xmark\\
	& cyc-inv & $0.1$ & \xmark\\
	& strongly-e & $-0.04$ & \cmark\\\hline

\multirow{3}*{
    \begin{tikzpicture}[scale=0.5]
    \foreach \i in {1, ..., 8} {
    	\draw (\i * 360 / 8:1) node[circle, fill=black, inner sep = 1pt] (n\i) {};
    };
    \draw (n1) -- (n2) -- (n3) -- (n4) -- (n5) -- (n6) -- (n7) -- (n8);
    \end{tikzpicture}
}	& perm-inv & $0.06$ & \cmark\\
	& cyc-inv & $0.05$ & \cmark\\
	& strongly-e & $-0.02$ & \xmark\\\hline
	
\multirow{3}*{
    \begin{tikzpicture}[scale=0.5]
    \foreach \i in {1, ..., 8} {
    	\draw (\i * 360 / 8:1) node[circle, fill=black, inner sep = 1pt] (n\i) {};
    };
    \draw (n1) -- (n2) -- (n3) -- (n4) -- (n5) -- (n6) -- (n7) (n4) -- (n6);
    \end{tikzpicture}
}	& perm-inv & $-0.14$ & \cmark\\
	& cyc-inv & $-0.03$ & \cmark\\
	& strongly-e & $0.1$ & \xmark\\\hline
	
\multirow{3}*{
    \begin{tikzpicture}[scale=0.5]
    \foreach \i in {1, ..., 8} {
    	\draw (\i * 360 / 8:1) node[circle, fill=black, inner sep = 1pt] (n\i) {};
    };
    \foreach \i in {1, ..., 7} {
    	\draw (n\i) -- (n8);
    };
    \end{tikzpicture}
}	& perm-inv & $0.16$ & \cmark\\
	& cyc-inv & $0.12$ & \cmark\\
	& strongly-e & $0.07$ & \cmark\\\hline
	
\multirow{3}*{
    \begin{tikzpicture}[scale=0.5]
    \foreach \i in {1, ..., 8} {
    	\draw (\i * 360 / 8:1) node[circle, fill=black, inner sep = 1pt] (n\i) {};
    };
    \draw (n8) -- (n1) -- (n2) (n1) -- (n3) (n8) -- (n7) -- (n6) (n7) -- (n5) (n8) -- (n4);
    \end{tikzpicture}
}	& perm-inv & $-0.03$ & \xmark\\
	& cyc-inv & $3\times 10^{-3}$ & \cmark\\
	& strongly-e & $-3\times 10^{-3}$ & \xmark\\\hline
\end{tabular}
    \label{tab:edge-case-results}
\end{table}

For the tests performed here, we have chosen a variety of special graphs that test various assumptions. In order, the first graph tests the aforementioned hypothesis that the neural network merely counts the number of edges. If this were the case, the prediction on the graph would indicate connectedness. From the single graph, it would therefore be easy to conclude that the hypothesis is refuted. However,  the label separation landscape is more complex, as seen from the subsequent graph of a complete graph of $n-1$ nodes linked with a single edge to the last node. The failure of  some of the neural networks to correctly classify the graph hints at another issue at play here, the difficulty of mapping discrete entities, graphs, onto a continuous space, the Hilbert space. In the embedding model of equation \eqref{eq:embedding}, similar graphs, as measured by the number of edges involved, are also mapped close to each other. As such, the decision boundary has to be quite sharp to correctly classify all cases. In the case of the first two graphs, the decision boundary is possibly not quite sharp enough.

This is also shown by the other examples. The third graph tests disconnected components with the same structure, which is apparently not well captured by the symmetric structures. 

Graphs four and five test the minimum connected and a same-sized disconnected graph, respectively. This is very close to the aforementioned threshold for connectedness mentioned above. It is likely that examples of this graph are part of the data set and this is reflected in the correct classification for the permutation-invariant and cyclic-invariant quantum circuits. 

The last two instances, graphs six and seven, test tree graphs with different depths. The second instance of a tree with depth two is more likely to be contained in the training data. The classification is correct in the first case and incorrect in the second case for the permutation-invariant case.

\paragraph{Strongly entangling layer} From the results presented in table \ref{tab:edge-case-results}, the performance of the strongly entangling layer is often in disagreement with the other circuits. This is due to the failure of the quantum circuit to converge to the data set, as can be gained from figure~\ref{fig:connectedness-result}. The results are effectively random and should not be construed as a reasonable result.

\paragraph{Accuracy} It should be noted that the score assigned to each graph also represents some form of confidence, on an arbitrary non-linear scale. Assuming $0.01$ as an arbitrary cut-off for a meaningful prediction, some of the presented results move from an incorrect classification to an undecided one. The permutation-invariant network generally provides a stronger signal, consistent with the overall performance on the validation set as shown in figure~\ref{fig:connectedness-result}. This links the averaged accuracy of the validation set with the particular performance on the individual edge cases. 

\section{Discussion}

In this work we have shown the applicability of edge case testing to graph classification and quantum machine learning. The method allows to refute some of the hypotheses that arise during the training of (quantum) neural networks. We suggest similar approaches in other cases as well, wherever the specific performance of a network could be reasonably explained by an easier model. In the present case, counting the number of edges in a graph is a good indicator of connectedness and the training data is not specifically tailored to provide cases where the simple model is refuted. Close examination of the quantum neural network performance beyond bare performance figures is a prerequisite to an eventual deployment.

\section*{Acknowledgments}

MBM acknowledges funding from the German Federal Ministry of Education and Research (BMBF) under the funding program ”Förderprogramm Quantentechnologien – von den Grundlagen zum Markt” (funding program quantum technologies – from basic research to market), project BAIQO, 13N16089.

\bibliographystyle{IEEEtranS}
\input{main.bbl}

\end{document}